\documentclass[]{jetp} 

\begin{document}
\English

\title{Electron Acceleration via Lower-Hybrid Drift Instability in Astrophysical Plasmas: Dependence on Plasma Beta and Suprathermal Electron Distributions}
\rtitle{Electron Acceleration via Lower-Hybrid Drift Instability in Astrophysical Plasmas\dots}

\author{Ji-Hoon}{Ha} 
\affiliation{Korea Space Weather Center, 63025, Jeju, South Korea}
\email{jhha223@korea.kr}

\author{Elena S.}{Volnova} 
\affiliation{Institute for Basic Science, 34126, Daejeon, South Korea}

\abstract{Density inhomogeneities are ubiquitous in space and astrophysical plasmas, particularly at magnetic reconnection sites, shock fronts, and within compressible turbulence. The gradients associated with these inhomogeneous plasma regions serve as free energy sources that can drive plasma instabilities, including the lower-hybrid drift instability (LHDI). Notably, lower-hybrid waves are frequently observed in magnetized space plasma environments, such as Earth's magnetotail and magnetopause. Previous studies have primarily focused on modeling particle acceleration via LHDI in these regions using a quasilinear approach. This study expands the investigation of LHDI to a broader range of environments, spanning weakly to strongly magnetized media, including interplanetary, interstellar, intergalactic, and intracluster plasmas. To explore the applicability of LHDI in various astrophysical settings, we employ two key parameters: (1) plasma magnetization, characterized by the plasma beta parameter, and (2) the spectral slope of suprathermal electrons following a power-law distribution. Using a quasilinear model, we determine the critical values of plasma beta and spectral slope that enable efficient electron acceleration via LHDI by comparing the rate of growth of instability and the damping rate of the resulting fluctuations. We further analyze the time evolution of the electron distribution function to confirm these critical conditions. Our results indicate that electron acceleration is generally most efficient in low-beta plasmas ($\beta < 1$). However, the presence of suprathermal electrons significantly enhances electron acceleration via LHDI, even in high-beta plasmas ($\beta > 1$). Finally, we discuss the astrophysical implications of our findings, highlighting the role of LHDI in electron acceleration across diverse plasma environments.}

\maketitle


\section{Introduction}
\label{sec:s1}
Particle acceleration through collisionless phenomena is ubiquitous in space and astrophysical plasmas. It is primarily facilitated by plasma waves generated from various instabilities, which are driven by regions containing free energy sources, such as shocks or associated velocity-space anisotropies [1-7]. The characteristics of these instabilities have been extensively examined using numerical methods, including particle-in-cell (PIC) simulations [2,3,5,6] and hybrid simulations [1,4,7]. Results from such simulations have highlighted that particle acceleration involves a variety of instabilities operating across multiple scales, from electron and ion kinetic scales to fluid scales. Building on the underlying physics of particle acceleration revealed by these simulation results, theoretical modeling has also been performed [8-14]. These models have significantly contributed to our understanding of in situ measurements in space plasma environments and the multi-wavelength radiation emitted by accelerated particles in galactic and extragalactic sources.

Inhomogeneities in magnetic field, velocity, density, and temperature, spanning fluid to kinetic scales, represent free energy sources that drive plasma instabilities in space and astrophysical plasmas. In particular, the diamagnetic drift associated with density gradients drives the lower-hybrid drift instability (LHDI) [15,16]. On electron kinetic scales, LHDI triggered at magnetic reconnection sites has been extensively studied [17-21]. Recent observations from the Magnetospheric Multiscale (MMS) mission have investigated the generation of lower-hybrid waves (LHWs) through LHDI at Earth's magnetopause [18,22-25]. Additionally, LHWs have been observed at plasma shock fronts, such as Earth's bow shock [26] and interplanetary shocks in the solar wind [27-29]]. Moreover, parallel electron heating induced by LHWs has been reported in Earth's magnetotail [30,31].

Along with the aforementioned observations, electron acceleration through LHDI in space plasma environments has also been studied using quasilinear models and numerical simulations [21,32-34]. The quasilinear models proposed in these studies describe how plasma waves with frequencies close to the lower-hybrid frequency transfer energy to particles through wave-particle interactions. While the quasilinear model does not account for the full plasma response to nonlinear interactions, it effectively describes the quasilinear growth stage of instability and the associated wave-particle interactions. In recent works [21,34], an extended quasilinear model has been proposed, which shows good agreement with results from full-kinetic simulations. Both the quasilinear model and full-kinetic simulations demonstrate that electron acceleration is most prominent during the quasilinear stage of LHDI, while the energy density of plasma waves driven by the instability saturates during the nonlinear stage of LHDI.

Considering the observational evidence, studies on LHDI and particle acceleration associated with LHWs have predominantly focused on space plasma environments. However, inhomogeneities are also expected to exist in various astrophysical media, including the interstellar and intracluster media. In the interstellar medium, local plasmas are likely to be inhomogeneous due to local sources such as pulsars [35,36], feedback from supernovae [37], and shocks associated with supernova remnants [38]. In the intracluster medium, evidence of inhomogeneous plasma is observed in the form of contact discontinuities in fluid dynamics, which indicate structures with opposing density and temperature gradients [39-42]. Numerical simulations of the intracluster medium further suggest that structures formed through gravitational collapse are inherently nonuniform [43,44]. The properties and time evolution of LHDI are expected to depend on the characteristics of the medium and the background particle distributions. In this context, LHDI likely plays a significant role in a variety of astrophysical environments beyond the near-Earth space.	
	
In this work, we adopt the quasilinear model for LHDI to specifically investigate electron acceleration and its nonlinear saturation in various astrophysical environments. Our analysis focuses on two major factors: (1) the properties of astrophysical media, ranging from weakly to strongly magnetized plasmas, and (2) the acceleration of suprathermal electron distribution functions, which are pre-accelerated by shocks or turbulence and deviate from a Maxwellian distribution. We anticipate that the results of this study will expand the applicability of LHDI across a broad range of systems, depending on the presence of suprathermal electrons and the amplitude of density gradients. Furthermore, exploring the role of LHDI in energy transport may enhance our understanding of the energy exchange between ions and electrons in turbulent media, a long-standing unsolved problem in astrophysical plasmas.
	
The organization of this paper is as follows. In Section 2, we describe the model framework for electron acceleration driven by LHDI. Section 3 presents the results of electron acceleration across a wide range of parameters, spanning weakly to strongly magnetized astrophysical plasmas. The implications of LHDI for collisionless thermal equilibration are also discussed. Finally, a brief summary is provided in Section 4.

\section{Model description}
\label{sec:s2}

In this work, we employ a quasilinear model to describe the wave-particle interaction between LHWs and electrons. The quasilinear theory is based on a second order perturbative expansion of the Vlasov equation, averaged over spatial variables. The model solves the following equations self-consistently [21,33,34]]:

\begin{equation}
\frac{\partial f_e(v_{\parallel},t)}{\partial t} 
= \frac{\partial}{\partial v_{\parallel}} 
\left[ D_e(v_{\parallel},t) \frac{\partial f_e(v_{\parallel},t)}{\partial v_{\parallel}} \right],
\end{equation}

\begin{equation}
D_e(v_{\parallel},t) 
= \frac{\pi e^2}{m_e^2} 
\int S_k(k_{\parallel},k_{\perp},t) 
\frac{k_{\parallel}^2}{k_{\perp}^2} 
\delta(\omega - k_{\parallel} v_{\parallel}) \, d^3k,
\end{equation}

\begin{equation}
\frac{\partial S_k(k_{\parallel},k_{\perp},t)}{\partial t}
= \left[ \Gamma_{\mathrm{LHDI}} \left( 1 - \frac{S_k}{S_{k,\max}} \right) 
+ \Gamma_e(k_{\parallel},k_{\perp},t) \right] S_k,
\end{equation}

\begin{equation}
\Gamma_{\mathrm{LHDI}} 
= \frac{\sqrt{2\pi}}{4} 
\frac{1}{\sqrt{1+\beta_i^2}}
\left( \frac{v_{Di}}{v_{thi}} \right)^2 \omega_{LH},
\end{equation}

\begin{equation}
\Gamma_e(k_{\parallel},k_{\perp},t) 
= \frac{\pi \omega_{LH}^2 \, \omega(k_{\parallel},k_{\perp})}{2 n_0 k_{\perp}^2} 
\frac{m_i}{m_e} 
\left. \frac{\partial f_e(v_{\parallel},t)}{\partial v_{\parallel}} 
\right|_{v_{\parallel}=\omega/k_{\parallel}}.
\end{equation}

Equation (1) describes the diffusion of the electron distribution function $f_e(v_{\parallel},t)$ in velocity space, driven by the diffusion coefficient $D_e(v_{\parallel},t)$, which is defined in Equation (2). This diffusion coefficient depends on the electric field energy density $S_k(k_{\parallel},k_{\perp},t)$, evolved by the LHDI and described in Equation (3). The evolution of $S_k$ reflects a competition between two physical processes. 

The term $\Gamma_{\mathrm{LHDI}}$ in Equation (4) represents the growth rate of LHDI. The growth rate term 
$\Gamma_{\mathrm{LHDI}} (1 - S_k / S_{k,\max})$ 
captures both the amplification of the wave energy and the saturation mechanism that limits the growth as $S_k$ approaches the theoretical maximum energy density $S_{k,\max}$. This saturation arises physically from the depletion of free energy in the ion drift or from nonlinear wave-wave interactions that inhibit further amplification. The validity of such saturation has been confirmed by fully kinetic simulations that show the saturation of LHDI following the initial growth stage [34]. Along with $\Gamma_{\mathrm{LHDI}}$, the growth of $S_k$ is also affected by the damping rate $\Gamma_e$ in Equation (5), which arises from wave-particle interactions with electrons. This term depends on the slope of the electron distribution function at the resonant velocity $v_{\parallel}=\omega/k_{\parallel}$. When the slope is negative (i.e., $\partial f_e(v_{\parallel},t)/\partial v_{\parallel}<0$), as is typical for Maxwellian-like distributions, electrons absorb energy from the wave via Landau damping, thereby reducing the overall growth rate. In such cases, the damping can prevent $S_k$ from reaching $S_{k,\max}$, resulting in a saturation level that is significantly lower than the theoretical maximum. The interplay between these two terms, $\Gamma_{\mathrm{LHDI}}$ and $\Gamma_e$, governs the saturation level and timescale of $S_k$. Within the quasilinear framework, this feedback mechanism results in a self-regulated steady-state wave intensity, dynamically determined by the evolving electron distribution.

The wave frequency $\omega(k_{\parallel},k_{\perp})$, and the wave damping rate 
$\Gamma_e(k_{\parallel},k_{\perp},t)$ and the spectrum 
$S_k(k_{\parallel},k_{\perp},t)$ are expressed in wavevector space 
$(k_{\parallel},k_{\perp})$, where $k_{\parallel}$ and $k_{\perp}$ are the wavenumbers 
parallel and perpendicular to the background magnetic field $B_0$, respectively. 
Additionally, the plasma characteristics are summarized as follows: (1) Plasma number density: $n_0$; (2) Lower-hybrid frequency: $\omega_{LH} = \sqrt{\omega_{ci}\,\omega_{ce}}$ where $\omega_{ci}$ and $\omega_{ce}$ are the ion and electron cyclotron frequencies, respectively; (3) Ion plasma beta: $\beta_i = {8\pi n_0 k_B T_i}/{B_0^2}$. Note that the magnetization of the plasma can also be parameterized by the ratio of the electron plasma frequency to the cyclotron frequency: ${\omega_{pe}^2}/{\omega_{ce}^2} = \beta_i^2 \left( {k_B T_i}/{m_i c^2} \right)^{-1}$. Throughout this paper, we use $k_B T_i = 1~\mathrm{keV}$.

When calculating the time evolution of $f_e(v_{\parallel},t)$ through Equations (1) - (5), we particularly consider the wavevector range of the LHDI fastest growing modes, 
which is $0.7 < k_{\perp} \rho_e < 1$ and $0 < k_{\parallel} \rho_i < 1$ [33], where $\rho_i$ and $\rho_e$ are the gyroradii of thermal ions 
and electrons, respectively. In this limited region of $k$-space, the wave spectrum is roughly 
approximated as $\omega(k_{\parallel},k_{\perp}) \simeq \omega_{LH}$. We employ the theoretical saturation energy density of the LHW wave electric field, 
as given by Lavorenti et al. [34]:
\begin{equation}
S_{\max} = \int S_{k,\max} \, d^3k = 
\begin{cases}
\dfrac{2 m_e}{m_i} \dfrac{(v_{Di}^2 / v_{thi}^2)}{1 + \omega_{pe}^2 / \omega_{ce}^2} \, 
n_0 k_B T_i, & \text{weak density gradient}, \\[1.2em]
\dfrac{2}{45\sqrt{\pi}} \dfrac{(v_{Di}^5 / v_{thi}^5)}{1 + \omega_{pe}^2 / \omega_{ce}^2} \, 
n_0 k_B T_i, & \text{strong density gradient}.
\end{cases}
\end{equation}
Assuming a homogeneous ion temperature $k_B T_i$, the diamagnetic drift velocity 
due to the ion density gradient is defined as
\begin{equation}
v_{Di} = \frac{ \mathbf{B}_0 \times (k_B T_i \nabla n_i)}{n_i e B_0^2}.
\end{equation}
The condition $v_{Di}/v_{thi} > 0.4$ corresponds to a strong density gradient, 
while smaller values indicate weaker density gradient.

\section{Results}
\label{sec:s3}

\subsection{Conditions for efficient electron acceleration through LHDI}
We derive the condition for efficient electron acceleration through LHDI based on parameters, including plasma beta and diamagnetic velocity. While it is necessary to solve the coupled Equations (1) - (5) to obtain the full spectral evolution through LHDI, we focus here on the very initial driving stage of LHDI, where the initial electron spectrum accelerated by wave-particle interaction is assumed to follow a power-law form:
\begin{equation}
f_e(v_{\parallel},t=0) = 
\frac{\eta_{\mathrm{spt}} n_0}{v_{the} A_{\kappa}} 
\left( \frac{v_{\parallel}}{v_{the}} \right)^{-\kappa}
= \frac{n_{\mathrm{spt}}}{v_{the} A_{\kappa}} 
\left( \frac{v_{\parallel}}{v_{the}} \right)^{-\kappa},
\end{equation}
with the normalization constant
\begin{equation}
A_{\kappa} \equiv \int_{v_{\parallel,\min}}^{v_{\parallel,\max}}
\left( \frac{v_{\parallel}}{v_{the}} \right)^{-\kappa} \frac{dv_{\parallel}}{v_{the}}
= \frac{1}{1-\kappa} \left[ 
\left( \frac{v_{\parallel,\max}}{v_{the}} \right)^{1-\kappa}
- \left( \frac{v_{\parallel,\min}}{v_{the}} \right)^{1-\kappa}
\right],
\end{equation}
where $A_{\kappa}$ is a dimensionless factor ensuring that $f_e$ is normalized to a number density per velocity.

The suprathermal fraction is then defined as
\begin{equation}
\eta_{\mathrm{spt}} = \frac{1}{n_0}
\int_{v_{\parallel,\min}}^{v_{\parallel,\max}} f_e(v_{\parallel},t=0)\, dv_{\parallel} 
= \frac{n_{\mathrm{spt}}}{n_0},
\end{equation}
where $n_{\mathrm{spt}} \equiv \eta_{\mathrm{spt}} n_0$ is the number density of suprathermal electrons defined by the fraction of suprathermal electrons $\eta_{\mathrm{spt}}$, and $\kappa$ is the spectral slope. When calculating $\eta_{\mathrm{spt}}$, the velocity range for wave-particle interaction $[v_{\parallel,\min}, v_{\parallel,\max}]$ is taken into account. Here, $v_{\parallel,\min}$ corresponds to the minimum velocity satisfying the condition of resonant wave-particle interaction, and $v_{\parallel,\max}$ is the upper cutoff defined by the maximum suprathermal tail. 

The derivatives of $f_e$ at $t=0$ are then expressed as follows:
\begin{equation}
\frac{\partial f_e(v_{\parallel},t=0)}{\partial v_{\parallel}}
= -\kappa v_{\parallel}^{-1} f_e(v_{\parallel},t=0),
\end{equation}
\begin{equation}
\frac{\partial^2 f_e(v_{\parallel},t=0)}{\partial v_{\parallel}^2}
= \kappa(\kappa+1) v_{\parallel}^{-2} f_e(v_{\parallel},t=0).
\end{equation}
The damping rate at the initial time, normalized to $\omega_{LH}$, is then
\begin{equation}
\frac{\Gamma_e(k_{\parallel},k_{\perp},t=0)}{\omega_{LH}}
\approx - \frac{\pi \omega_{LH}^2}{2 n_0 k_{\perp}^2 v_{the}} 
\frac{m_i}{m_e} \, 
\kappa \left( \frac{\omega_{LH}}{k_{\parallel} v_{the}} \right)^{-1} 
f_e(v_{\parallel},t=0).
\end{equation}
Here, the damping rate is expressed using the velocity normalized by the electron thermal velocity $v_{the}$. Importantly, $\Gamma_e(k_{\parallel},k_{\perp},t=0)$ is evaluated at the initial time of the quasilinear stage, and its value depends explicitly on the suprathermal fraction and the spectral slope $\kappa$. While Fig. 1 and Fig. 2 show $\Gamma_e(k_{\parallel},k_{\perp},t=0)$ for the electrons with $v_{\parallel} = v_{the}$ as horizontal reference lines for comparison with the LHDI growth rates, the full time-dependent evolution of $\Gamma_e(k_{\parallel},k_{\perp},t)$ follows the evolving electron distribution self-consistently in the quasilinear framework, which is shown in the following section.

Fig. 1 shows the growth rate of LHDI as a function of $v_{Di}$. We expect that electrons are efficiently accelerated through LHDI when the growth rate of LHDI exceeds the damping rate. For instance, in the case with $\beta_i = 1$ and $\kappa = 10$, efficient acceleration is expected when $v_{Di}/v_{thi} > 0.2$. The minimum diamagnetic velocity required for efficient acceleration increases as $\beta_i$ increases. Additionally, electron acceleration becomes inefficient when considering a steeper initial electron distribution function, as a steeper spectral slope enhances the damping process.

Fig. 2 shows the growth rate of LHDI as a function of $\beta_i$. In the case of a weak density gradient ($v_{Di}/v_{thi} = 0.2$; panel (a)), the maximum value of $\beta_i$ for $\kappa = 10$ is roughly $\beta_i \sim 3$. This indicates that the growth of LHDI is unlikely in high-$\beta_i$ environments, such as the intracluster medium ($\beta_i \sim 50$--$100$). In such environments, the growth of LHDI is only possible in the presence of preaccelerated particles through shock or turbulence (i.e., the initial electron distribution function with a flat spectral slope). In the case of a strong density gradient ($v_{Di}/v_{thi} = 0.5$; panel (b)), on the other hand, the growth of LHDI is favorable over a wide range of $\beta_i$ ($\beta_i \lesssim 200$), regardless of the slope of the particle distribution function.

We next examine the time evolution of the electric field energy density associated with the LHDI, focusing on how the growth and damping rates influence its evolution. Fig. 3 presents the time evolution of $S_k$ for three different cases, parameterized by $\beta_i$ and $v_{Di}/v_{thi}$. To investigate the impact of the initial suprathermal population, we analyze the evolution of $S_k$ for different $\kappa$ values, ranging from 2 to 10. In all three cases shown in Fig. 3, the LHDI growth timescale decreases as $\kappa$ decreases. When the particle distribution has a higher $\kappa$ (closer to Maxwellian), rapid damping suppresses the LHDI growth, reducing the energy available for particle acceleration compared to distributions with lower $\kappa$. This indicates that energy transfer from LHDI-driven fluctuations to particles is more efficient in systems with stronger suprathermal populations 
(lower $\kappa$). According to the cases shown in Fig. 3, the influence of the suprathermal population becomes more prominent under conditions of weaker density gradients or higher $\beta_i$. Particularly, in panel (b), the effect of $\kappa$ is less significant due to the strong density gradient in a low-$\beta_i$ plasma. These characteristics observed in the time evolution of the LHDI are expected to influence electron acceleration by LHDI-driven fluctuations.

We then examine the wave-particle interaction at the initial quasilinear growth stage of LHDI. At this stage, we compute the diffusion coefficient and acceleration timescale by considering a short perturbative time interval $\delta t \ll t$, during which the distribution function remains close to its initial state. To extract a physically intuitive estimate, we reduce Equation (3) using the scaling $k_{\parallel} \sim \mathcal{O}(\rho_i^{-1})$ and $k_{\perp} \sim \mathcal{O}(\rho_e^{-1})$, corresponding to typical LHDI wave numbers. Under this approximation, the spectral integral becomes:
\begin{equation}
\int S_k(k_{\parallel},k_{\perp},\delta t) \frac{k_{\parallel}^2}{k_{\perp}^2} 
\delta(\omega - k_{\parallel} v_{\parallel}) \, d^3 k 
\sim \mathcal{O}\left(\frac{\rho_i^{-2}}{\rho_e^{-2}}\right) 
S_0 \exp\left[ (\Gamma_{\mathrm{LHDI}} + \Gamma_e(\mathcal{O}(\rho_i^{-1}), \mathcal{O}(\rho_e^{-1}), t=0)) \, \delta t \right],
\end{equation}
where $S_0$ is the initial electric energy density. The exponential term $\exp\left[ (\Gamma_{\mathrm{LHDI}} + \Gamma_e) \delta t \right]$, which describes the growth and damping of LHDI during the perturbed timescale, is derived under the assumption of a short time interval during the early quasilinear growth stage, such that $S_k(\delta t) \sim S_0 \exp\left[ (\Gamma_{\mathrm{LHDI}} + \Gamma_e) \delta t \right]$. This is consistent with a linearized approximation of Equation (3) when $S_k \ll S_{k,\max}$. 

Adopting Equation (14), the diffusion coefficient around $t \approx 0$ is roughly derived as:
\begin{equation}
D_e(v_{\parallel},0) \approx D_e(v_{\parallel},\delta t) 
= \frac{\pi e^2}{m_e^2} \int S_k(k_{\parallel},k_{\perp},\delta t) 
\frac{k_{\parallel}^2}{k_{\perp}^2} \delta(\omega - k_{\parallel} v_{\parallel}) \, d^3 k 
\propto \exp\left[ (\Gamma_{\mathrm{LHDI}} + \Gamma_e(\mathcal{O}(\rho_i^{-1}), \mathcal{O}(\rho_e^{-1}), t=0)) \, \delta t \right].
\end{equation}
Physically, this proportionality reflects that, during the initial quasilinear growth stage, the rate at which electrons diffuse in velocity space is directly controlled by the balance between the LHDI growth rate $\Gamma_{\mathrm{LHDI}}$ and the electron damping rate $\Gamma_e(\mathcal{O}(\rho_i^{-1}), \mathcal{O}(\rho_e^{-1}), t=0)$. A higher LHDI growth rate enhances diffusion and accelerates electrons, whereas stronger damping reduces the diffusion, inhibiting acceleration. This expression therefore provides a physically meaningful, order-of-magnitude estimate for the diffusion coefficient during the early stage of LHDI and serves as a reference point in analyzing particle acceleration efficiency. 

Using the diffusion coefficient around $t \approx 0$, we define the dimensionless pseudo-diffusion coefficient as follows:
\begin{equation}
D^*_{e,0} / (\omega_{pe} v_{the}^2) \sim 
\exp\left[ (\Gamma_{\mathrm{LHDI}} + \Gamma_e(\mathcal{O}(\rho_i^{-1}), \mathcal{O}(\rho_e^{-1}), t=0)) \, \delta t \right].
\end{equation}
The pseudo-acceleration timescale, using the pseudo-diffusion coefficient, is then calculated as:
\begin{equation}
\omega_{pe} \tau_{\mathrm{acc}}(v) = \frac{(v/v_{the})^2}{D^*_{e,0}/(\omega_{pe} v_{the}^2)}.
\end{equation}
The pseudo-acceleration timescales for different sets of parameters are shown in Fig. 4 Panel (a) presents the effect of $\beta_i$. The acceleration timescale increases rapidly as $\beta_i$ exceeds 10. Additionally, as shown in panel (b), acceleration becomes slower with a steeper spectral slope, as the damping rate is proportional to the spectral slope. Panels (c) and (d) display the results showing how the acceleration timescales depend on the diamagnetic velocity (or the density gradient of the system). In particular, the effect of the diamagnetic drift is more pronounced for the case of an initial distribution function with a steeper slope.

Based on the analysis performed above, we define the critical plasma beta and diamagnetic velocity ($\beta_{\mathrm{crit}}$ and $(v_{Di}/v_{thi})_{\mathrm{crit}}$) for efficient electron acceleration through LHDI, as shown in Fig. 5. Here, $\beta_{\mathrm{crit}}$ and $(v_{Di}/v_{thi})_{\mathrm{crit}}$ represent the maximum $\beta_i$, and the minimum $v_{Di}$ required for efficient electron acceleration. It is shown that $\beta_{\mathrm{crit}}$ increases as $v_{Di}/v_{thi}$ increases, and $(v_{Di}/v_{thi})_{\mathrm{crit}}$ decreases as $\beta_i$ decreases. This reflects that the acceleration efficiency could be enhanced for systems with smaller $\beta_i$ and larger $v_{Di}/v_{thi}$. While the specific values of the critical $\beta_i$ and $v_{Di}$ depend on the free parameter $\eta_{\mathrm{spt}}$, the dependence of the growth rate of LHDI on $\beta_i$, $v_{Di}$, and $\kappa$ is independent of the value of $\eta_{\mathrm{spt}}$.

\subsection{Time evolution of the electron distribution function during the quasilinear growth stage of LHDI}

Adopting the initial electron distribution function described in Equation~(7), we solve the set of Equations (1) - (5) self-consistently. In this section, we fully consider the effects of plasma beta, diamagnetic velocity, and the characteristics of the initial distribution function, which are parameterized by the spectral slope, in the self-consistently evolving system. Additionally, while the monochromatic wave with 
$k_{\parallel} \sim \mathcal{O}(\rho_i^{-1})$ and $k_{\perp} \sim \mathcal{O}(\rho_e^{-1})$, satisfying the resonant condition $\omega_{LH} - k_{\parallel} v_{\parallel} \simeq 0$, was considered in the previous section for simplicity, here we expand our consideration to the region of wavevector space corresponding to the fastest growing modes of LHDI for wave-particle interaction, that is, $0.7 < k_{\perp} \rho_e < 1$ and $0 < k_{\parallel} \rho_i < 1$ [33].

Panels (a) - (c) of Fig. 6 show the time evolution of the electron distribution function for different values of $\beta_i$, ranging from 0.1 to 100. For example, in the case of $\kappa = 10$, electron acceleration becomes less efficient as $\beta_i$ increases, which is consistent with the finding shown in Fig. 5 (a). This indicates that systems with $\beta_i > 10$ rapidly saturate due to the balance between growth and damping, and thus the results obtained from the self-consistently evolving system align with those obtained from the initial distribution function alone. It is shown that while acceleration in systems with $\beta_i = 10$ and 100 is enhanced when considering flatter spectra (i.e., $\kappa = 2$ and 5), the acceleration efficiency decreases as $\beta_i$ increases for spectral slopes ranging from 2 to 10. Additionally, panels (d) - (f) of Fig. 6 illustrate the effects of diamagnetic velocity on the time evolution of the electron distribution function. In particular, the initial electron distribution with a steeper slope (i.e., $\kappa = 10$) is efficiently accelerated only in the presence of a strong density gradient ($v_{Di}/v_{thi} > 0.4$). In contrast, for $\kappa = 2$, electron acceleration is observed in both weak and strong density gradient regimes. This finding is also consistent with the results shown in Fig. 5 (b), which demonstrate that the minimum diamagnetic velocity required for electron acceleration decreases as the initial particle distribution becomes flatter (i.e., $\kappa$ decreases).

We further examine the electron energy fraction obtained through LHDI. The electron kinetic energy of the given distribution function $f_e(v_{\parallel},t)$ is expressed as:
\begin{equation}
E_e(t) = \int_{v_{\parallel} > v_{the}} \frac{1}{2} m_e v_{\parallel}^2 f_e(v_{\parallel},t) \, dv_{\parallel}.
\end{equation}
Fig. 7 shows the time evolution of the electron kinetic energy normalized by the initial electron kinetic energy $E_{e,0} = E_e(t=0)$. We investigate the dependence of electron acceleration on $\beta_i$ and $v_{Di}/v_{thi}$ for different values of the suprathermal index $\kappa$, ranging from 2 to 10. As also seen in the electron distribution functions presented in Fig. 6, the fraction of accelerated electron energy increases as $\kappa$ decreases, indicating that suprathermal populations enhance wave-particle interaction. The effects of $\beta_i$ and $v_{Di}/v_{thi}$ are clearly observed across the range of $\kappa$ values. In the cases of $\beta_i < 1$, electron acceleration increases over time as a consequence of the modification of the damping rate due to the time evolution of the electron distribution. Considering the energy budget of LHDI relative to the background thermal energy, this $\beta_i$-dependence is reasonable, as the electromagnetic energy in the lower-$\beta_i$ 
system becomes more significant compared to the higher-$\beta_i$ system. For systems with $\beta_i > 10$, however, the system rapidly saturates due to the balance between growth and damping of LHDI (see panels (a) - (c)). In this high-$\beta_i$ regime, electron acceleration is more likely to occur in systems with a strong density gradient ($v_{Di}/v_{thi} > 0.4$) as shown in panels (d) - (f).

\subsection{Nonlinear saturation of LHDI and the effects of suprathermal electron population}
We employ an extended quasilinear model, as proposed in previous works [21,34], to investigate the nonlinear saturation stage of LHDI beyond the quasilinear stage. This extended model incorporates nonlinear physics, as demonstrated by full-kinetic numerical simulations. It has been shown that electron acceleration occurs prior to the nonlinear saturation of LHDI, within a characteristic nonlinear timescale denoted as $\tau_{\mathrm{NL}}$. The extended model is formulated as follows:
\begin{equation}
\frac{\partial f_e(v_{\parallel},t)}{\partial t} = 
\frac{\partial}{\partial v_{\parallel}} \left[ D_{\mathrm{NL}}(v_{\parallel},t) 
\frac{\partial f_e(v_{\parallel},t)}{\partial v_{\parallel}} \right],
\end{equation}
where the nonlinear diffusion coefficient $D_{\mathrm{NL}}(v_{\parallel},t)$ is defined as:
\begin{equation}
D_{\mathrm{NL}}(v_{\parallel},t) =
\begin{cases} 
D_e(v_{\parallel},t), & t < \tau_{\mathrm{NL}} \\[2mm]
0, & t \ge \tau_{\mathrm{NL}}
\end{cases}.
\end{equation}
The characteristic nonlinear timescale $\tau_{\mathrm{NL}}$ is expressed as:
\begin{equation}
\tau_{\mathrm{NL}} \sim \frac{3}{2} \, \omega_{ci}^{-1} 
\left(\frac{m_i}{m_e}\right)^{1/2} 
\left(\frac{V_{Di}}{V_{thi}}\right)^{-2} 
\left(1 + \frac{T_e}{T_i}\right)^{-1}.
\end{equation}
For simplicity, we have assumed initial thermal equilibrium, $T_e \simeq T_i$. However, as electrons are accelerated, $T_e$ may increase and consequently modify $\tau_{\mathrm{NL}}$ through the factor $(1+T_e/T_i)^{-1}$. A more quantitative assessment of this feedback requires dedicated kinetic simulations and is left for future work.

We specifically examine how plasma properties, such as plasma beta and the spectral form of the particle distribution, influence the nonlinear saturation of the energy density produced by LHDI. Fig. 8 presents the saturated energy density ($S_{\mathrm{NL,sat}}$) across weakly to strongly magnetized plasmas. The results show that the saturated energy density decreases as the density gradient weakens, consistent with findings from previous studies [21,34]. In the low-$\beta_i$ regime, the saturated energy density in systems with weak density gradients is highly sensitive to the spectral slope of the initial electron distribution (panel (a)). In contrast, such dependence is not observed in systems with strong density gradients (panel (b)). However, even in the presence of a strong density gradient, a dependence on the spectral slope is evident. Notably, in the high-$\beta_i$ regime, the saturated energy density decreases with steeper spectral slopes. These findings demonstrate that the initial plasma conditions significantly influence the dynamical evolution of LHDI during both its quasilinear and nonlinear phases.

\section{Summary and discussion}
\label{sec:s4}
This study aims to expand the applicability of the quasilinear model for particle acceleration through LHDI from space plasma environments to various astrophysical media, ranging from weakly to strongly magnetized plasmas. Specifically, we examine how the magnetization of plasma, diamagnetic velocity, and the spectral form of the initial particle distribution affect the growth and damping of LHDI during the quasilinear growth stage and the associated particle acceleration. In the absence of sufficient suprathermal electrons (i.e., cases with steeper slopes, such as $\kappa = 10$), particle acceleration through LHDI is significant only when $\beta_i < 1$. This finding aligns with the fact that the majority of previous works have focused on space plasma environments with $\beta_i < 1$ for particle acceleration through LHDI [32 - 34]. On the other hand, when considering a sufficient fraction of suprathermal electrons (i.e., cases such as $\kappa = 2$ or $\kappa = 5$), the acceleration efficiency in the higher-beta environments (e.g., $\beta_i \sim 10$ - 100) increases compared to that for an initial electron distribution with a steeper spectral slope. Additionally, strong diamagnetic drift generally enhances the acceleration efficiency through LHDI. Regarding the evolution of LHDI, even in the strongly nonlinear phase, the effect of diamagnetic drift significantly influences the saturated energy density of LHDI. In systems with strong density gradients, the possible range of plasma beta facilitating energy transfer through LHDI is extended.

The astrophysical application of LHDI found in this work is summarized as follows. In low-beta environments, the acceleration efficiency may deviate from the estimates based on the initial Maxwellian distribution. Plasma properties and local phenomena, such as shocks and turbulence, vary significantly across different media, including the interplanetary and interstellar medium. For instance, the interstellar medium, which contains strong shocks driven by supernova remnants, is expected to have a larger fraction of suprathermal particles compared to the interplanetary medium, where shocks are relatively weak. This is supported by numerical simulations showing that acceleration efficiency at shocks increases with increasing shock Mach number [45]. In this regard, the acceleration efficiency through LHDI is expected to be significantly influenced by the physical conditions of astrophysical media in low-beta regimes. Additionally, LHDI could also serve as a possible mechanism for energy transfer in high-beta plasmas, since inhomogeneous density and the presence of suprathermal electrons are likely, at least locally, in high-beta environments such as the intracluster medium [11,43,44].

Before closing, we further discuss the future applicability of LHDI in astrophysical problems. It has been demonstrated that thermal disequilibration between ions and electrons can be induced by various mechanisms, including turbulence cascades 46,47], particle acceleration in reconnection sites [48-50] such as Earth's magnetotail [51-53], shocks in supernova remnants [54-56], heliospheric shocks [57,58], and shocks in galaxy clusters [59]. In particular, ions are preferentially heated by plasma turbulence in plasmas with $\beta_i>1$, whereas electron heating dominates in plasmas with $\beta_i<1$ [46]. The mechanism responsible for achieving thermal equilibrium remains a long-standing problem in astrophysical plasmas. For instance, in the intracluster medium, which is a representative example of high-beta plasma, collisional relaxation between ions and electrons is unlikely, as the ion-electron relaxation timescale is comparable to the dynamical timescale of the intracluster medium [60,61]. This highlights the necessity for collisionless mechanisms to facilitate thermal equilibrium. Given the generation mechanism of LHDI, the free energy in the ion distribution (e.g., density gradients and diamagnetic drift) can be transferred to the electron distribution. In this regard, LHDI may play a role in establishing thermal equilibrium. According to the results of this study, collisionless equilibration through LHDI could be further enhanced in regions where electrons are pre-accelerated, such as in shocks or turbulent environments.



\section*{Acknowledgements}
We thank to anonymous referees for providing constructive comments to improve the manuscript.


\begin{figure}[t]
    \includegraphics[width=\textwidth]{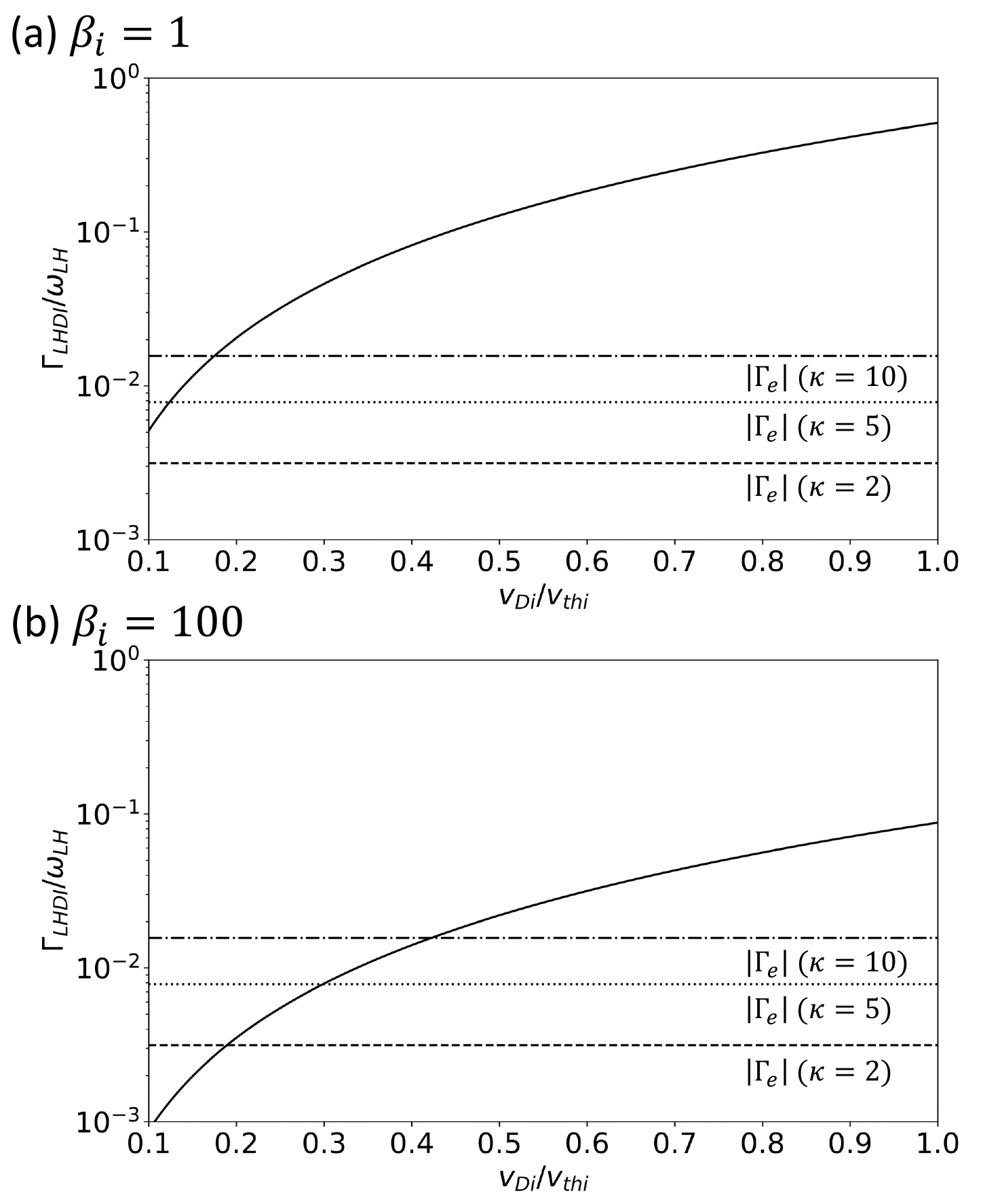}
    \caption{Growth rates of LHDI for $\beta_i=1$ and $\beta_i=100$ as functions of $v_{Di}/v_{thi}$. For comparison, the dashed, dotted, and dash-dotted horizontal lines represent the damping rates $\Gamma_e$ for the electrons with $v_{\parallel}=v_{the}$ evaluated at $t=0$ for spectral slopes $\kappa = 2$, $\kappa = 5$, and $\kappa = 10$, respectively. The suprathermal fraction is set to $\eta_{\rm spt} = 10^{-3}$.}
    \label{fig:f1}
\end{figure}

\begin{figure}[t]
    \includegraphics[width=\textwidth]{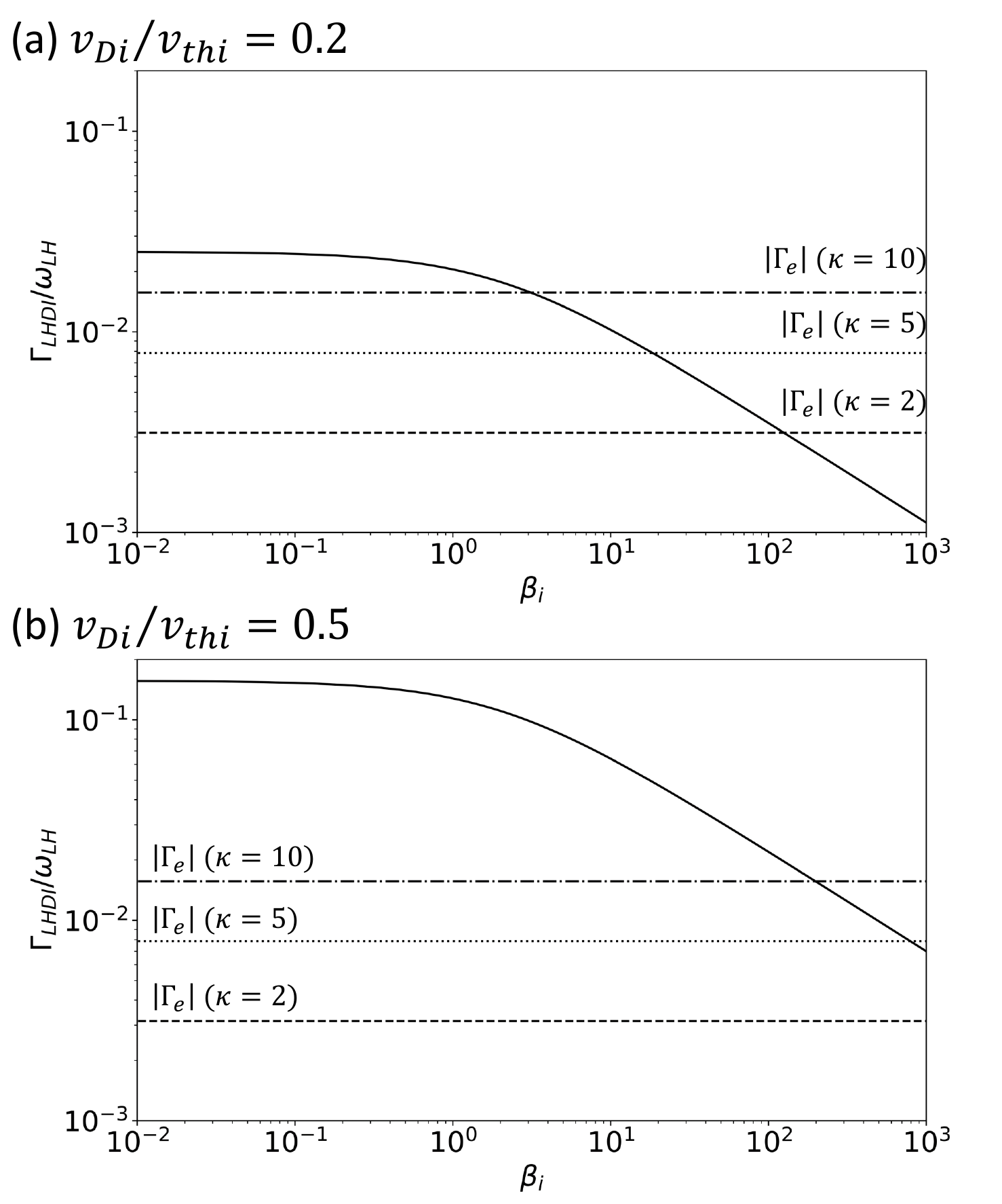}
    \caption{Growth rates of LHDI for $v_{Di}/v_{thi} = 0.2$ and $v_{Di}/v_{thi} = 0.5$ as functions of $\beta_i$. For comparison, the dashed, dotted, and dash-dotted horizontal lines represent the damping rates $\Gamma_e$ for the electrons with $v_{\parallel}=v_{the}$ evaluated at $t=0$ for spectral slopes $\kappa = 2$, $\kappa = 5$, and $\kappa = 10$, respectively. The suprathermal fraction is set to $\eta_{\rm spt} = 10^{-3}$.}
    \label{fig:f2}
\end{figure}

\begin{figure}[t]
    \includegraphics[width=0.7\textwidth]{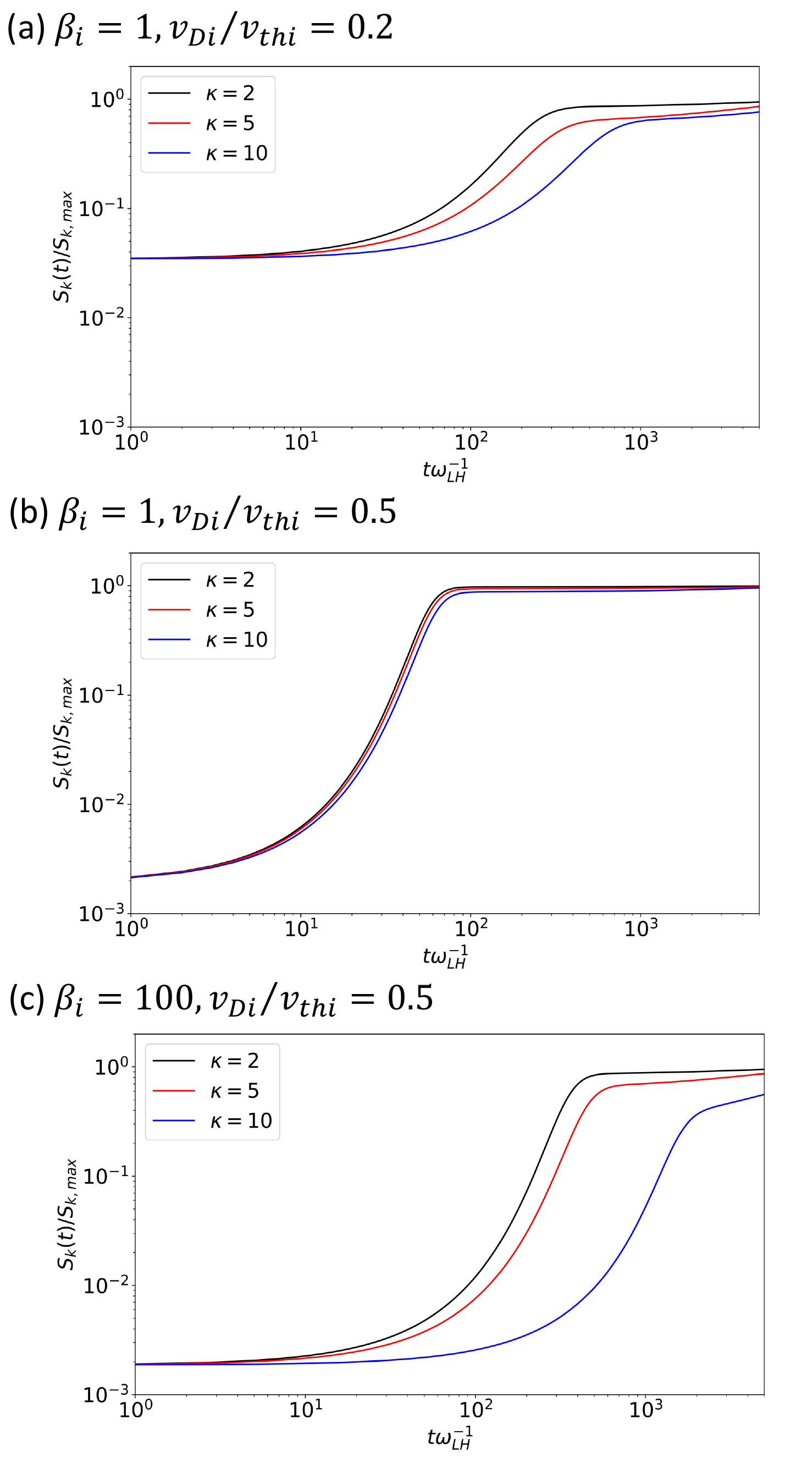}
    \caption{Time evolution of the electric field energy density ($S_k$) for three parameter sets: (a) $\beta_i = 1, \, v_{Di}/v_{thi} = 0.2$, (b) $\beta_i = 1, \, v_{Di}/v_{thi} = 0.5$, and (c) $\beta_i = 100, \, v_{Di}/v_{thi} = 0.5$.}
    \label{fig:f3}
\end{figure}

\begin{figure}[t]
    \includegraphics[width=\textwidth]{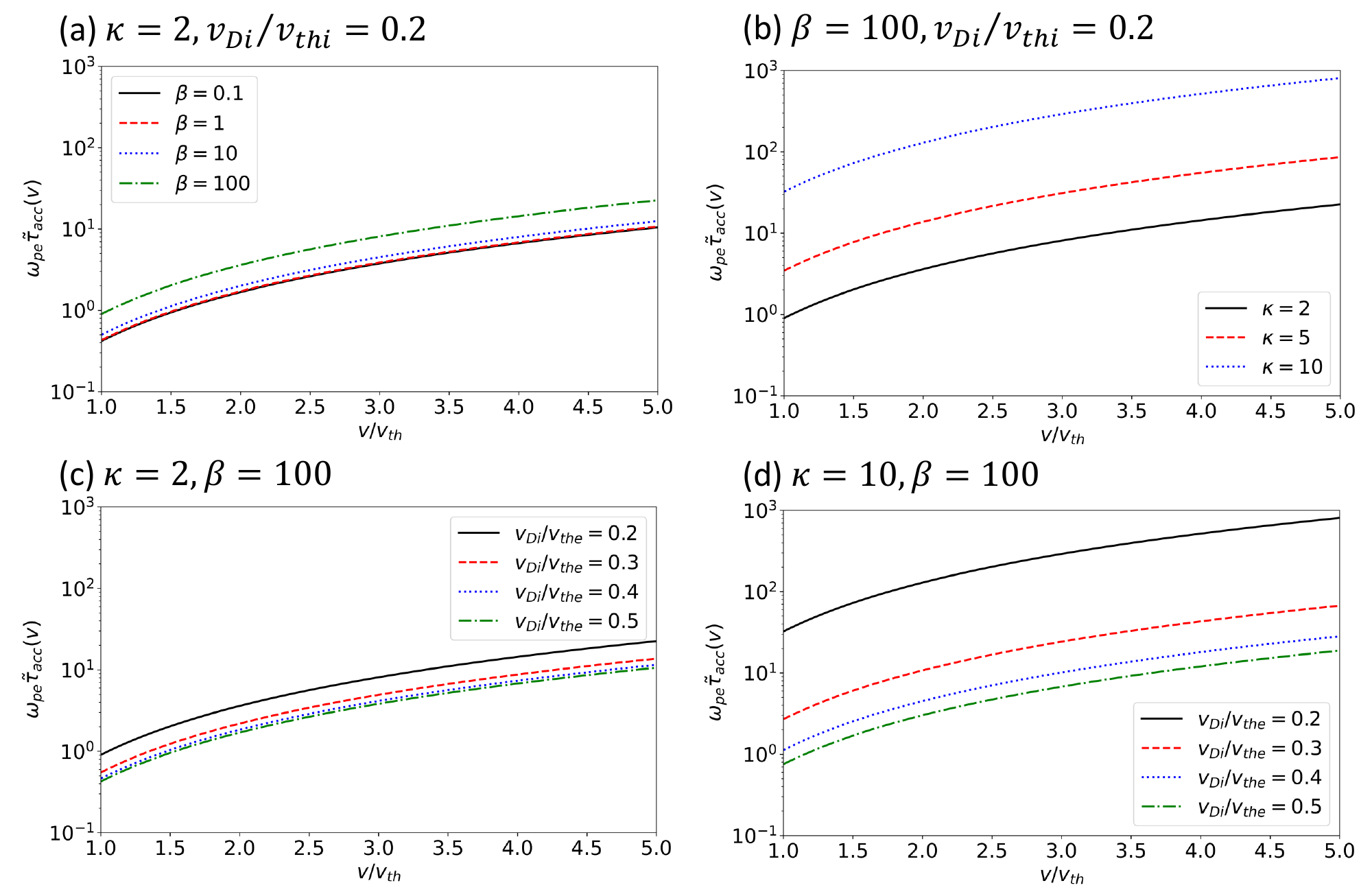}
    \caption{Pseudo-acceleration timescales for different sets of parameters, i.e., $\kappa$, $\beta_i$, and $v_{Di}$.}
    \label{fig:f4}
\end{figure}

\begin{figure}[t]
    \includegraphics[width=\textwidth]{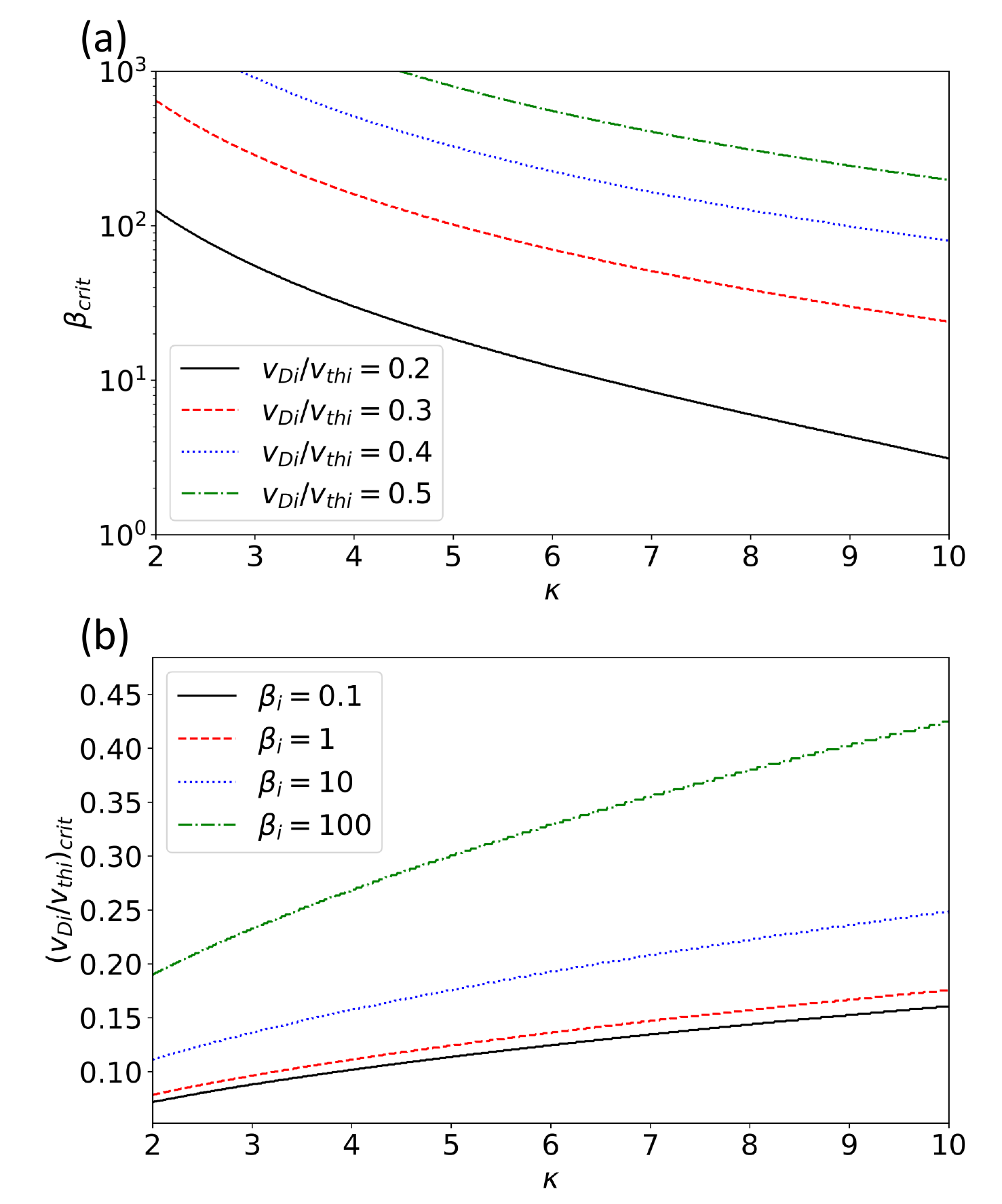}
    \caption{Critical beta (panel (a)) and diamagnetic velocity (panel (b)) for efficient electron acceleration through LHDI as functions of the slope $\kappa$. $\eta_{\rm spt} = 10^{-3}$ is used.}
    \label{fig:f5}
\end{figure}

\begin{figure}[t]
    \includegraphics[width=\textwidth]{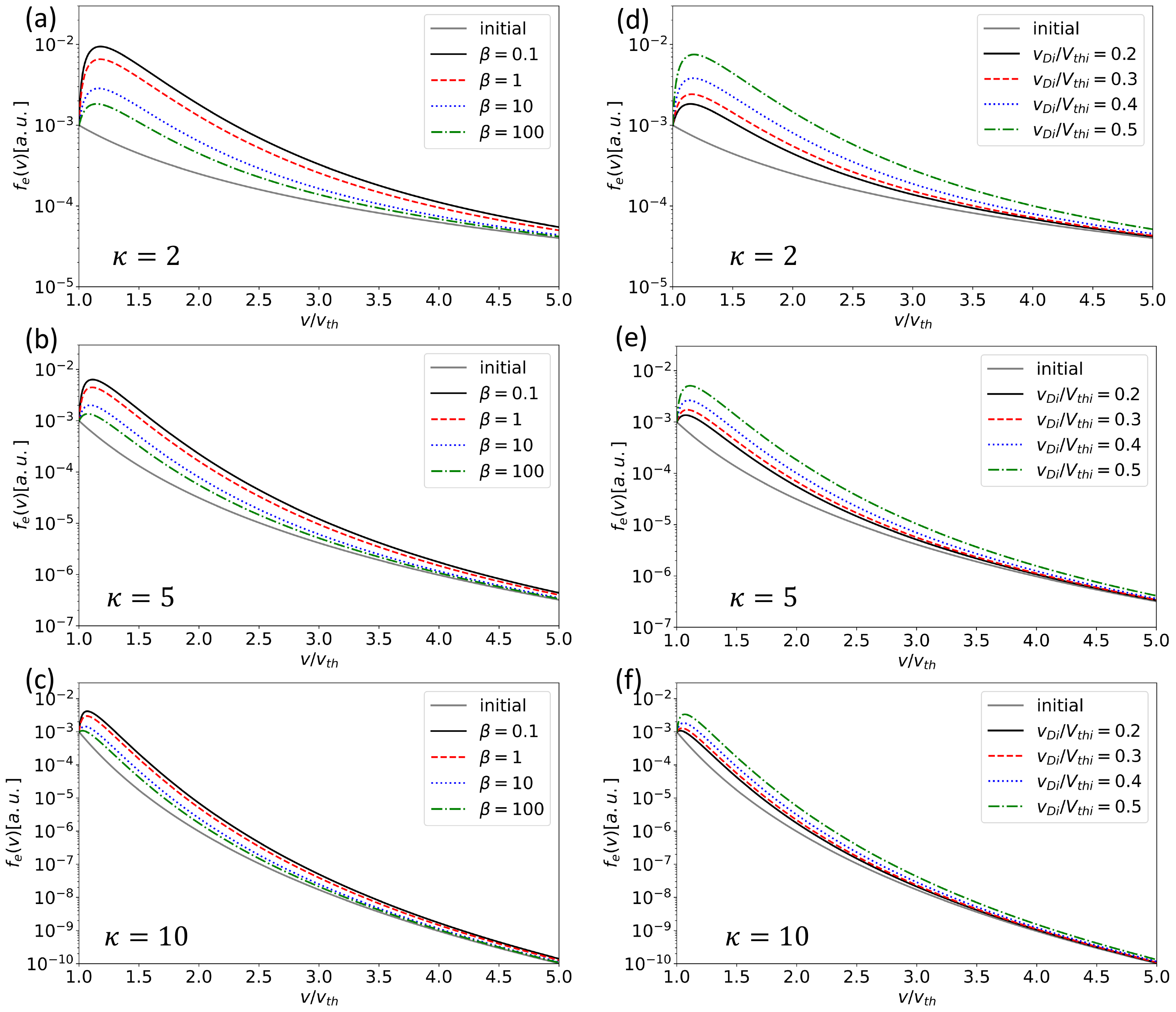}
    \caption{Electron distribution functions measured at $\omega_{LH} t = 100$. The dependence on plasma beta is shown in panels (a) - (c), while the dependence on diamagnetic velocity is shown in panels (d) - (f). The initial electron distribution function is displayed as a gray solid line, with spectral slopes $\kappa$ ranging from 2 to 10, and the suprathermal fraction set to $\eta_{\rm spt} = 10^{-3}$. $v_{Di}/v_{thi} = 0.2$ is used for panels (a) - (c), and $\beta_i = 100$ is used for panels (d) - (f).}
    \label{fig:f6}
\end{figure}

\begin{figure}[t]
    \includegraphics[width=\textwidth]{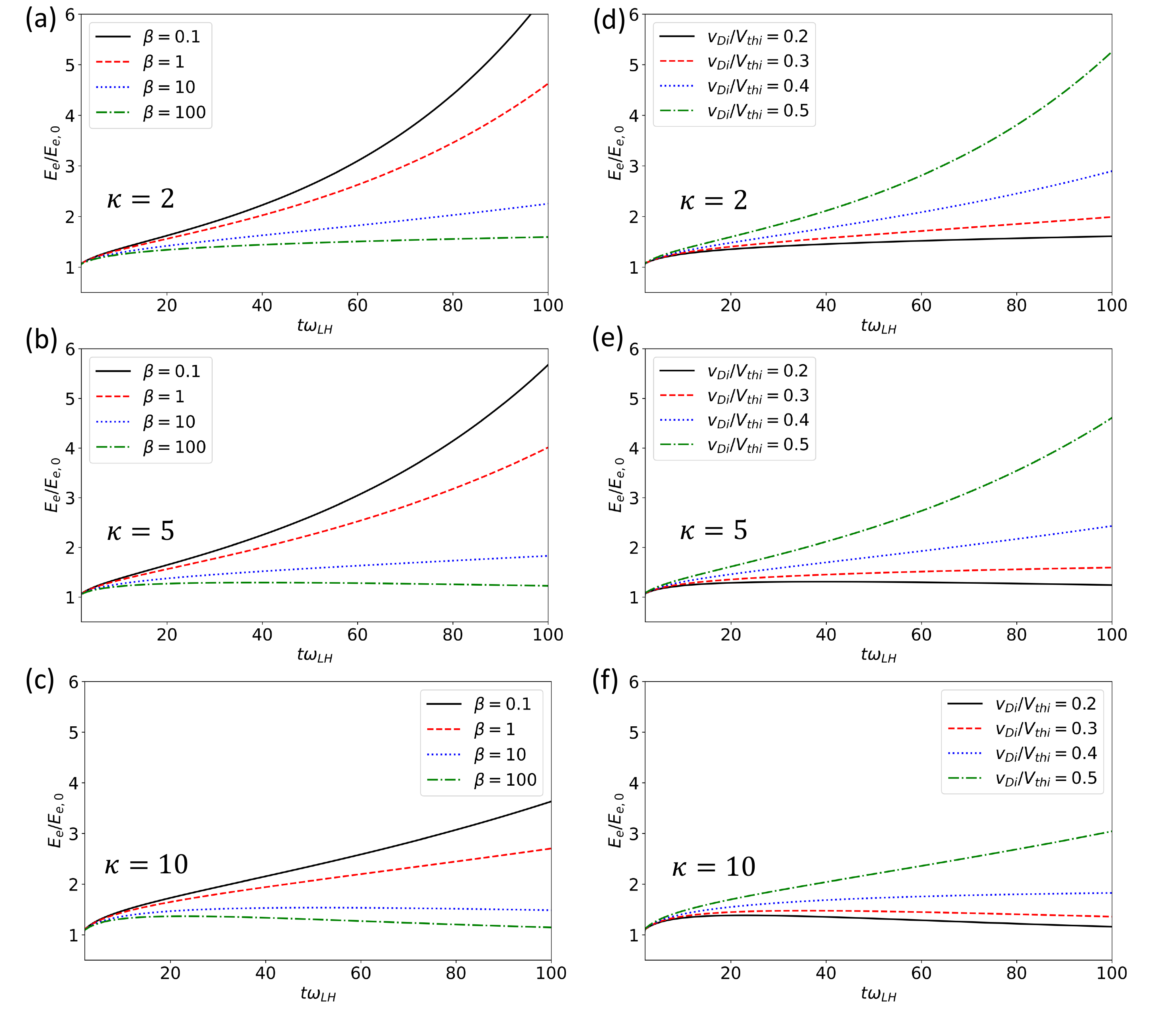}
    \caption{Time evolution of electron energy resulting from acceleration by LHDI. Panels (a) - (c) show the dependence on $\beta_i$, while panels (d) - (f) show the dependence on $v_{Di}/v_{thi}$. For panels (a) - (c), $v_{Di}/v_{thi} = 0.2$ is used, and for panels (d) - (f), $\beta_i = 100$ is used.}
    \label{fig:f7}
\end{figure}

\begin{figure}[t]
    \includegraphics[width=\textwidth]{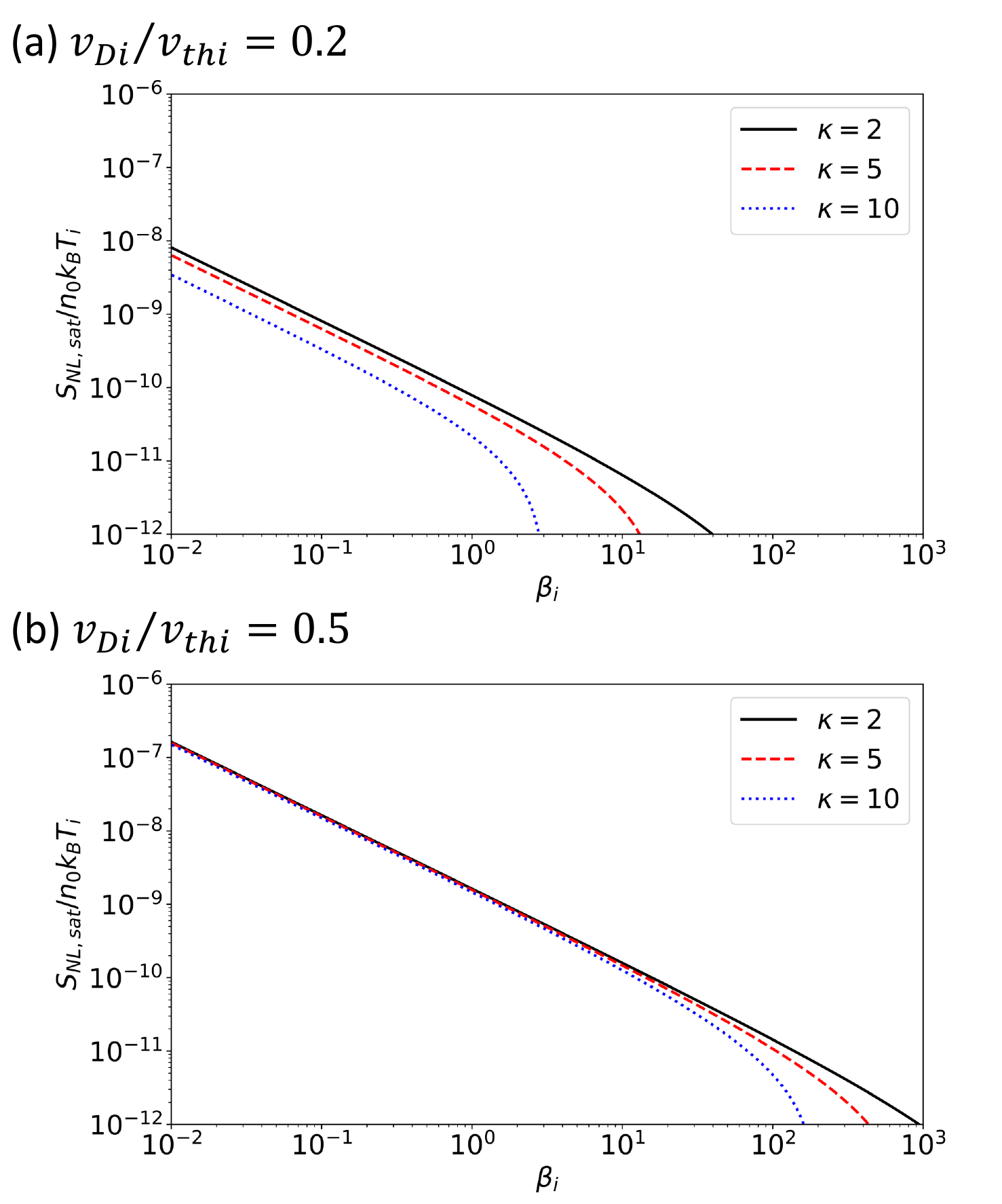}
    \caption{The saturated energy density in the nonlinear phase as a function of plasma beta for the cases of weak (panel (a)) and strong (panel (b)) density gradients. Initial spectra with spectral slopes $\kappa$ ranging from 2 to 10 are used.}
    \label{fig:f8}
\end{figure}


\begin{references}
\bibitem{1}
J. Giacalone, {\it The Astrophysical Journal}, 609, 452 (2004).

\bibitem{2}
M. Scholer and D. Burgess, {\it Physics of Plasmas}, 14, 072103 (2007).

\bibitem{3}
T. Umeda, Y. Kidani, S. Matsukiyo and R. Yamazaki, {\it Journal of Geophysical Research: Space Physics}, 117, A03206 (2012).

\bibitem{4}
D. Caprioli and A. Spitkovsky, {\it The Astrophysical Journal}, 794, 46 (2014).

\bibitem{5}
S. Kim, J.-H. Ha, D. Ryu and H. Kang, {\it The Astrophysical Journal}, 892, 85 (2020).

\bibitem{6}
S. Kim, J.-H. Ha, D. Ryu and H. Kang, {\it The Astrophysical Journal}, 913, 35 (2021).

\bibitem{7}
L. Orusa and D. Caprioli, {\it Physical Review Letters}, 131, 095201 (2023).

\bibitem{8}
R. Vainio, L. Kocharov and T. Laitinen, {\it The Astrophysical Journal}, 528, 1015 (2000).

\bibitem{9}
V.N. Zirakashvili, V.S. Ptuskin, and V\"{o}lk, H.J., {\it The Astrophysical Journal}, 678, 255 (2008).

\bibitem{10}
A. M. Bykov, D. C. Ellison, S. M. Osipov and A. E. Vladimirov, {\it The Astrophysical Journal}, 789, 137 (2014).

\bibitem{11}
J.-H. Ha, D. Ryu and H. Kang, {\it The Astrophysical Journal}, 907, 26 (2021).

\bibitem{12}
J.-H. Ha, {\it Astrophysics}, 67, 330 (2024a).

\bibitem{13}
J.-H. Ha, {\it Astrophysics and Space Science}, 369, 126 (2024b).

\bibitem{14}
J.-H. Ha, {\it Zhurnal Experimentalnoi i Teoreticheskoi Fiziki}, 167, 129 (2025).

\bibitem{15}
N. A. Krall and P. C. Liewer, {\it Physical Review A}, 4, 2094 (1971).

\bibitem{16}
R. C. Davidson, N. T. Gladd, C. S. Wu and J. D. Huba, {\it Physics of Fluids}, 20, 301 (1977).

\bibitem{17}
S. D. Bale, F. S. Mozer and T. Phan, {\it Geophysical Research Letters}, 29, 2180 (2002).

\bibitem{18}
D. B. Graham, Y. V. Khotyaintsev, C. Norgren, et al., {\it Journal of Geophysical Research: Space Physics}, 124, 8727 (2019).

\bibitem{19}
J. Yoo, J.-Y. Ji, M. V. Ambat, et al., {\it Geophysical Research Letters}, 47, e87192 (2020).

\bibitem{20}
J. Ng, J. Yoo, L. J. Chen, N. Bessho and H. Ji, {\it Physics of Plasmas}, 30, 042101 (2023).

\bibitem{21}
Y. Ren, L. Dai, C. Wang and Z. Guo, {\it The Astrophysical Journal}, 956, 143 (2023).

\bibitem{22}
D. B. Graham, Y. V. Khotyaintsev, C. Norgren, et al., {\it Journal of Geophysical Research: Space Physics}, 122, 517 (2017).

\bibitem{23}
M. Zhou, J. Berchem, R. J. Walker, et al., {\it Journal of Geophysical Research: Space Physics}, 123, 1834 (2018).

\bibitem{24}
B.-B. Tang, W. Y. Li, D. B. Graham, et al., {\it Geophysical Research Letters}, 47, e89880 (2020).

\bibitem{25}
Y. Ren, L. Dai, C. Wang and B. Lavraud, {\it The Astrophysical Journal}, 928, 5 (2022).

\bibitem{26}
S. N. Walker, M. A. Balikhin, H. S. C. K. Alleyne, et al., {\it Annales Geophysicae}, 26, 699 (2008).

\bibitem{27}
V. V. Krasnoselskikh, E. N. Kruchina, A. S. Volokitin and G. Thejappa, {\it Astronomy and Astrophysics}, 149, 323 (1985).

\bibitem{28}
Y. Zhang and H. Matsumoto, {\it Journal of Geophysical Research: Space Physics}, 103, 20561 (1998).

\bibitem{29}
L. B. Wilson, A. Koval, A. Szabo, et al., {\it Journal of Geophysical Research: Space Physics}, 118, 5 (2013).

\bibitem{30}
Y. V. Khotyaintsev, C. M. Cully, A. Vaivads, M. Andre and C. J. Owen, {\it Physical Review Letters}, 106, 165001 (2011).

\bibitem{31}
D.-X. Pan, Y. V. Khotyaintsev, D. B. Graham, et al., {\it Geophysical Research Letters}, 45, 116 (2018).

\bibitem{32}
K. G. McClements, R. Bingham, J. J. Su, J. M. Dawson and D. S. Spicer, {\it The Astrophysical Journal}, 409, 465 (1993).

\bibitem{33}
I. H. Cairns and B. F. McMillan, {\it Physics of Plasmas}, 12, 102110 (2005).

\bibitem{34}
F. Lavorenti, P. Henri, F. Califano, S. Aizawa and N. Andre, {\it Astronomy and Astrophysics}, 652, A20 (2021).

\bibitem{35}
E. N. Fadeev, A. S. Andrianov, M. S. Burgin, et al., {\it Monthly Notices of the Royal Astronomical Society}, 480, 4199 (2018).

\bibitem{36}
M. V. Popov and T. V. Smirnova, {\it Astronomy Reports}, 65, 1129 (2021).

\bibitem{37}
D. Martizzi, C.-A. Faucher-Giguere and E. Quataert, {\it Monthly Notices of the Royal Astronomical Society}, 450, 504 (2015).

\bibitem{38}
C. F. McKee, {\it Proceedings of IAU Colloquium 101, Cambridge University Press}, 205 (1988).

\bibitem{39}
M. Markevitch, T. J. Ponman, P. E. J. Nulsen, et al., {\it The Astrophysical Journal}, 541, 542 (2000).

\bibitem{40}
M. Markevitch and A. Vikhlinin, {\it Physics Reports}, 443, 1 (2007).

\bibitem{41}
H. Bourdin, P. Mazzotta, M. Markevitch, S. Giacintucci and G. Brunetti, {\it The Astrophysical Journal}, 764, 82 (2013).

\bibitem{42}
J. ZuHone and E. Roediger, {\it Journal of Plasma Physics}, 82, 535820301 (2016).

\bibitem{43}
J.-H. Ha, D. Ryu and H. Kang, {\it The Astrophysical Journal}, 857, 26 (2018).

\bibitem{44}
S. Roh, D. Ryu, H. Kang, S. Ha and H. Jang, {\it The Astrophysical Journal}, 883, 138 (2019).

\bibitem{45}
R. Xu, D. Caprioli and A. Spitkovsky, {\it The Astrophysical Journal Letters}, 897, L41 (2020).

\bibitem{46}
Y. Kawazura, M. Barnes, A. A. Schekochihin, et al., {\it Proceedings of the National Academy of Sciences}, 116, 771 (2019).

\bibitem{47}
J. Squire, R. Meyrand and M. W. Kunz, {\it The Astrophysical Journal Letters}, 957, L30 (2023).

\bibitem{48}
M. A. Shay, C. C. Haggerty, T. D. Phan, et al., {\it Physics of Plasmas}, 21, 122902 (2014).

\bibitem{49}
C. C. Haggerty, M. A. Shay, J. F. Drake, et al., {\it Geophysical Research Letters}, 42, 9657 (2015).

\bibitem{50}
M. Hoshino, {\it The Astrophysical Journal Letters}, 868, L18 (2018).

\bibitem{51}
W. Baumjohann, G. Paschmann and C. A. Cattell, {\it Journal of Geophysical Research}, 94, 6597 (1989).

\bibitem{52}
C.-P. Wang, M. Gkioulidou, L. R. Lyons and V. Angelopoulos, {\it Journal of Geophysical Research: Space Physics}, 117, A08215 (2012).

\bibitem{53}
J. P. Eastwood, T. D. Phan, J. F. Drake, et al., {\it Physical Review Letters}, 110, 225001 (2013).

\bibitem{54}
C. E. Rakowski, {\it Advances in Space Research}, 35, 1017 (2005).

\bibitem{55}
J. C. Raymond and K. E. Korreck, {\it AIP Conference Proceedings}, 781, 342 (2005).

\bibitem{56}
P. Ghavamian, J. M. Laming and C. E. Rakowski, {\it The Astrophysical Journal}, 654, L69 (2007).

\bibitem{57}
S. J. Schwartz, M. F. Thomsen, S. J. Bame and J. Stansberry, {\it Journal of Geophysical Research}, 93, 12923 (1988).

\bibitem{58}
C. T. Russell, {\it AIP Conference Proceedings}, 781, 3 (2005).

\bibitem{59}
H. R. Russell, B. R. McNamara, J. S. Sanders, et al., {\it Monthly Notices of the Royal Astronomical Society}, 423, 236 (2012).

\bibitem{60}
S. Ettori and A. C. Fabian, {\it Monthly Notices of the Royal Astronomical Society}, 293, L33 (1998).

\bibitem{61}
M. Takizawa, {\it The Astrophysical Journal}, 509, 579 (1998).


\end{references}
\end{document}